# Title: Nanophotonic control of circular dipole emission: towards a scalable solid-state to flying-qubits interface

**Authors:** B. le Feber[1], N. Rotenberg[1], L. Kuipers[1*]

**Affiliations:**

[1]FOM Institute AMOLF, Science Park 104, 1098 XG Amsterdam, The Netherlands

*Correspondence to: kuipers@amolf.nl

**Abstract:**

Controlling photon emission by single quantum emitters with nanostructures is crucial for scalable on-chip quantum information processing. Nowadays nanoresonators can affect the lifetime of emitters and ultimately induce strong coupling between the emitters and the light field, while nanoantennas can control the directionality of the emission. Expanding this control to the manipulation of the emission of orbital angular momentum-changing transitions would enable coupling between long-lived solid-state qubits and flying qubits. As these transitions are associated with circular rather than linear dipoles, such control requires detailed knowledge of the spatially dependent interaction of a complex dipole with highly structured optical eigenstates containing local helicity. Using a classical analogue, we experimentally map the coupling of circular dipoles to photonic modes in a model structure, a photonic crystal waveguide. We show that depending on the local helicity the dipoles can be made to couple to modes either propagating to the left or to the right. The maps are in excellent agreement with calculations. Our measurements, therefore, demonstrate the coupling of spin to photonic pathway with near-unity (0.8 ± 0.1) efficiency.

**Main text**

Control of the emission properties of circular dipole sources, where the phase of the orthogonal linear dipole components cannot be neglected, with a scalable nanophotonic interface would constitute a tremendous step towards viable, on-chip quantum information processing. This control would allow for manipulation of the emission properties associated with the long-lived spin states of solid-state emitters, such as quantum dots[1,2,3,4] and nitrogen vacancy centers[5,6,7], as their orbital angular momentum-changing decay to specific spin states is associated with the helicity of *circular* transition dipoles[8,9] (Fig. 1). Furthermore, if we interface nanophotonic structures with spin qubits, all the lessons learned from the study of how such structures interact with linear dipoles to, for example, control their decay rates[10,11,12] or the directionality of their emission[13,14] could be immediately applied. With such an interface we could couple, or even entangle, solid-state emitters to photonic pathways, essentially encoding the quantum information of a long lived solid-state qubit onto a versatile flying qubit[15,16], allowing for a new avenue towards quantum information processing elements .

Clearly, tremendous benefits can be obtained from a controlled interface of nanophotonics with spin-states of emitters. Ideally, such an interface will be deterministic, meaning that all emission is into the desired modes of the nanophotonic structure and not into free space, and that each distinct spin state of the emitter is coupled to a single photonic pathway. The former requirement can be met by, for example, placing a quantum emitter inside a photonic crystal waveguide, which enables the extraction of over 98 percent of the QD emission[17]. The latter requirement has, to date, not been demonstrated, although researchers have recently shown preferential emission of QDs situated at the crossing of two ridge waveguides[9]. It was shown that, depending on the helicity of its circular transition dipole, a QD preferentially emits into two of the four exit waveguides.

For a deterministic interface of emitter spin to photon pathway, detailed knowledge of the projection of a circular dipole onto highly structured optical eigenstates of a nanophotonic structure, which contain local helicity[18], is required. Such knowledge is key if the emitter is to be correctly positioned. In practice, the fine details of the optical modes are highly dependent on the geometry of the nanostructure, and hence are sensitive to imperfections. Moreover, fabricating emitters in precise locations on nanophotonic structures is a complex and challenging procedure, which ultimately imposes additional constraints on the feasibility of a solid-state to flying-qubit interface. For example, the interaction of the nanophotonic structure and the emitter must be relatively constant over an area defined by the precision with which the emitter can be placed, typically on the order of tens of nanometers[19,20]. Consequently, a demonstration of viable nanophotonic interface for solid-state and flying qubits must fulfill two requirements. First, efficient and directional coupling between a circular dipole and a photon pathway must be observed on a real nanophotonic structure. Second, a full spatial mapping of the interaction of the dipole and this structure must be created.

Here, we use a classical, tunable dipole source to demonstrate near-perfect coupling of helicity to photon pathway in a silicon photonic crystal waveguide (PhCW). First, we experimentally show that the radiation from the tip of a near-field optical microscope probe can mimic that of a linear transition dipole. Subsequently, we extend this method to emulate the emission of a circular transition dipole, i.e. one that is associated with a change in the emitter's spin state. By scanning this tunable source we create high resolution spatial maps of its emission into the PhCW, for different circular dipoles. We show that the helicity of the light emitted by such circular dipoles in combination with the unique local helicity of the photonic eigenstates of the PhCW leads to efficient and deterministic directional emission. We underpin these experimental observations with a rigorous theoretical framework that describes the radiation of circular dipoles near a PhCW. Because the emission of a photon by a transition dipole and a classical dipole into a PhCW is identical in the weak coupling limit, our result demonstrates that scalable spin-to-pathway coupling is possible.

**Emission control with photonic crystal waveguides**

The way in which an emitter, and in particular a quantum emitter, radiates when placed near a nanophotonic structure is a subject of intense research [10,11,12,13,14,21,22,23]. The decay of a two-level system, an inherently quantum process, is associated with a transition dipole (Fig. 1a,b), which arises due to charge redistribution that occurs in the emitter during the transition. The radiation of the transition dipole is identical to that of a classical dipole, except that a transition dipole only exists for the duration of a single radiation event. When an emitter is placed near a nanophotonic structure the radiation probability of the transition dipole is altered by the number of photonic states into which it can emit. If the nanophotonic structure is a PhCW, and the emitter is located in the air above its surface, the factor by which emission is changed relative to free-space, is (Supplementary Section I).

$$F_{L,R}(\mathbf{d},\mathbf{r},\omega) = \frac{3\pi c^2 a n_g(\omega)}{2\omega^2} \left|\hat{\mathbf{d}}^* \cdot \mathbf{e}_{L,R}(\mathbf{r},\omega)\right|^2, \tag{1}$$

where $\mathbf{e}_{L,R}$ are the electric (magnetic) fields of the normalized left- and right-propagating modes of the PhCW that interact with the dipoles of orientation $\hat{\mathbf{d}}$, which can be electric ($\hat{\mathbf{p}}$) or magnetic ($\hat{\mathbf{m}}$) in nature, or both. Here, $\mathbf{r}$ the emitter position, $\omega$ the emission frequency, $a$ is the PhCW lattice period, $n_g$ is the group index, and the subscripts $L$ and $R$ explicitly show that $F$ can depend on the direction (in our case: left or right) that light propagates in the PhCW. Lastly, we note that this expression assumes that all emission is into the photonic crystal waveguide modes[24]. Due to the structural symmetry of PhCWs, $\mathbf{e}_L(\mathbf{r},\omega) = \mathbf{e}_R^*(\mathbf{r},\omega)$. From equation (1) it is clear that for perfectly directional emission into the PhCW we need $|\hat{\mathbf{d}}^* \cdot \mathbf{e}_L(\mathbf{r}_0,\omega)| = 1$ and $|\hat{\mathbf{d}}^* \cdot \mathbf{e}_R(\mathbf{r}_0,\omega)| = 0$, or vice versa, for certain positions $\mathbf{r}_0$. For circular dipoles to maximally directionally emit into the PhCW, therefore, we require positions where the left and right propagating modes are circularly polarized, but with the opposite helicities.

**Modification of linear dipole emission**

Photonic crystal waveguides, by virtue of their highly structured near-field distributions that locally sample all in-plane vectorial field orientations[18], seem ideally suited to couple emission to pathway. Moreover, PhCWs can also slow down light [25], enhancing emission when $n_g$ is large (equation (1), Fig. 2a). This enhancement can be observed in the calculated $F_{L,R}(\hat{\mathbf{p}}_{x,y})$ presented in Fig. 2b (see Methods), for linear $\hat{\mathbf{p}}_x$ and $\hat{\mathbf{p}}_y$ emitting at the wavelengths marked in the dispersion relation (Fig. 2a). Similar $F_{L,R}(\hat{\mathbf{m}}_{x,y})$ maps quantifying the interaction of the PhCW with magnetic dipoles can be created (Supplementary Section II). For $F_{L,R}(\hat{\mathbf{p}}_{x,y})$, the maximum enhancement factor changes from 1.6 at 1575 nm ($n_g = 15$) to 9.8 at 1592 nm nm ($n_g = 120$). Furthermore, we observe that with increasing $n_g$ high enhanced emission factors become available away from $y = 0$ (dashed blue line Fig. 2b). This is expected since the PhCW modes typically spread out as the light slows down[25,26]. Importantly, for all $n_g$, the maps of the linear dipole emission modification factor $F_R(\hat{\mathbf{p}}_{x,y})$ (not shown) are identical to those of $F_L(\hat{\mathbf{p}}_{x,y})$. So for a linear dipole the emission is not directional.

We mimic the behavior of a dipolar emitter using the emission from a 210 nm wide, near-field scanning optical microscope (NSOM) probe, which is known to act as a sort of continuous transition dipole[27,28,29] (Fig. 2d). We raster scan the probe in the sample plane at a height of 20 nm. While scanning, we collect the light coupled to the waveguide in a heterodyne detection scheme that gives us access to its phase (Fig. 2c). We use the phase information to Fourier filter unwanted reflections from the waveguide end facets[18], mapping out the excitation efficiencies of both the left- and rightwards propagating modes, $D_L(\hat{\mathbf{d}})$ and $D_R(\hat{\mathbf{d}})$, respectively.

**Helicity to pathway coupling**

Because we wish to couple helicity, or spin, to path, we need to confirm that we can control the orientation and phase of our dipole source. We start by showing that we can create any linear dipole ($\lambda_{freespace} = 1575$ nm), in this case by varying the orientation of our dipole from $\hat{\mathbf{x}}$ to $\hat{\mathbf{y}}$. At this wavelength the PhCW interacts with magnetic dipoles in much the same way as it does with electric dipoles (Supplementary Section II). Consequently, we limit the following discussion to electric dipoles. We vary the orientation of the dipole by controlling the polarization of the light that we inject into our near-field probe, and show the measured $D_{L,R}(\hat{\mathbf{p}}_x)$ and $D_{L,R}(\hat{\mathbf{p}}_y)$ in Fig. 3a. These maps show that $x$- and $y$- oriented dipoles emit in drastically different ways into the PhCW, but as expected we do not observe directional emission. These measurements are in excellent agreement with the calculations (Fig. 2b, and line cuts in Fig. 3a).

To create circular dipoles, such as those associated with orbital angular momentum-changing transitions ($m_s = \pm 1$ states, Fig. 1b), we need to control the phase of our dipole source. Hence, we add a liquid crystal variable wave plate to our polarization control optics (LC in Fig. 2c, for details see Supplementary Section IV), which allows us to create $\mathbf{d} = (\hat{\mathbf{p}}_x \pm i\hat{\mathbf{p}}_y)/\sqrt{2}$ sources. Strikingly, both the measured $D_{L,R}$ and the calculated $F_{L,R}$ for these circular sources (Fig. 3b) now exhibit a pronounced asymmetry. In fact, we observe that flipping either the emission direction ($D_L$ or $D_R$) or dipole helicity ($(\hat{\mathbf{p}}_x + i\hat{\mathbf{p}}_y)/\sqrt{2}$ or $(\hat{\mathbf{p}}_x - i\hat{\mathbf{p}}_y)/\sqrt{2}$) results in a mirroring of the excitation map about the middle of the waveguide ($y = 0$, dashed blue lines Fig. 3b). For example, the emission of a right-handed dipole (top row of panels Fig. 3b) is predominantly to the left when the dipole is placed over the bottom half of the waveguide ($y < 0$), and to the right when the dipole is located in the top half ($y > 0$). We take line cuts (along the dashed white lines in Fig. 3b) to illustrate the high degree of coupling between helicity and direction. For both dipole handednesses maximal emission in one direction corresponds to a minimum in the other. Furthermore, in both experiment and calculations we find positions where the emitted photons from a circular dipole source are coupled to a direction with unitary probability (Supplementary Section V).

For our approach to dipole helicity to pathway coupling to be a viable route towards on-chip quantum technology, not only the directionality but also the emission rate itself needs to be maximized. Consequently, we define the *directional coupling efficiency*, $\eta$, to be

$$\eta(\mathbf{r},\omega) = \frac{\left[D_L(\hat{\mathbf{d}}_{LCP}) - D_R(\hat{\mathbf{d}}_{LCP})\right] - \left[D_L(\hat{\mathbf{d}}_{RCP}) - D_R(\hat{\mathbf{d}}_{RCP})\right]}{2D_{max}}, \qquad (2)$$

where the subscripts denote propagation direction ($L$, $R$) or handedness ($LCP$, $RCP$) and $D_{max}$ is the maximum intensity found in any of $D_L(\hat{\mathbf{d}}_{LCP})$, $D_R(\hat{\mathbf{d}}_{LCP})$, $D_L(\hat{\mathbf{d}}_{RCP})$ and $D_R(\hat{\mathbf{d}}_{RCP})$. If $\eta(\mathbf{r},\omega)=0$, either the chance of emitting a photon left and right is equal, or no photons are emitted at $\mathbf{r}$. Conversely, if $|\eta(\mathbf{r},\omega)|=1$, then at $\mathbf{r}$ spin is *both* deterministically coupled to path *and* a circular dipole emits maximally.

Fig. 4a (first panel) presents $\eta(\mathbf{r},\omega)$ determined for emission at $1575\,\text{nm}$, corresponding to the individually measured emission maps shown in Fig. 3b. Astonishingly, we observe relatively large regions where the helicity of the dipole almost perfectly determines the pathway that a photon will take, both to the left ($\eta \to 1$, red regions) and to the right ($\eta \to -1$, blue regions). In fact, we observe a maximal helicity-to-pathway coupling of $|\eta|_{max}^{exp} = 0.8 \pm 0.1$, where the deviation from unity is largely due to experimental noise. Altogether, these measurements confirm that PhCWs are ideally suited to act as an interface between flying and solid-state qubits that conforms with the current state-of-the-art quantum emitter placement capabilities [19,20].

We underpin our observations by calculating the theoretical efficiencies, using equation (1) and (2) with $\hat{\mathbf{d}} = (\hat{\mathbf{p}}_x \pm i\hat{\mathbf{p}}_y)/\sqrt{2}$. The calculations (second panel of Fig. 4a) are in excellent agreement with the measurements, as can also be seen from the line cuts (right two panels) taken at positions of maximal directionality. The maximum theoretical directional coupling efficiency ($|\eta|_{max}^{calc}$) for this PhCW is $0.95$, meaning that at this wavelength a circular dipole has the highest and most directional emission at the same locations. Because of the excellent agreement between experiment and theory, we can extend the calculations to the center height of the PhCW slab. At the center, where radiation to free space is almost completely suppressed[17], we also predict near-unity helicity to path coupling efficiency (Supplementary Section II, VII).

Finally, for completeness, we note that at this wavelength the calculated $\eta$ for both electric and magnetic dipoles are almost identical (Fig. S6), with both peaking near unity. This ideal helicity-to-pathway coupling efficiency, for both electric and magnetic dipoles, again highlights the potential of PhCWs in on-chip quantum applications.

**Tuning the coupling efficiency**

The optical properties of PhCWs can be tuned through geometry [30]. To explore this while using a single structure, we vary the excitation wavelength from $1575\,\text{nm}$ to $1585\,\text{nm}$ and $1592\,\text{nm}$, with calculated $n_g$'s of, 40 and 120, respectively. For our PhCW, the modes, and hence $F$ (Fig. 2b), at these wavelengths spread out away from the center of the waveguide. Strikingly, in both the measurements and calculations at these higher $n_g$'s, regions of highly directional coupling appear away from the center of the waveguide (Figs. 4b ($n_g = 40$) and c ($n_g = 120$), Supplementary Section V). Further, as can be observed from the line cuts in Fig. 4, as $n_g$ increases, the position of $|\eta|_{max}$ shifts from the central region of the waveguide, to the area of the holes (and from the green to the red dashed line). Importantly, we observe that $\eta_{max}$ decreases as $n_g$ increases, from $0.8 \pm 0.1$ at $n_g = 15$, to $0.6 \pm 0.1$ at $n_g = 40$, and $0.7 \pm 0.1$ at $n_g = 120$. This decrease in directional coupling efficiency is due to the combined emission of electric and magnetic dipoles that do not share identical $\eta$ profiles (Supplementary Section III, V), and because for the higher $n_g$'s, the maxima in directionality and emission amplitude are no longer located at the same position. It is, however, relatively simple to design a PhCW where $|\eta|_{max} = 1$ over a broad range of $n_g$'s (Supplementary Section VI). That is, for a point-like quantum emitter that emits into a propagating mode of the PhCW, perfect spin-to-pathway coupling is, in principle, always possible.

**Conclusions**

In conclusion, we have shown through classical measurements that a PhCW can be used to deterministically couple circular dipole helicity to pathway. Experimentally, we observe a coupling efficiency of $0.8 \pm 0.1$, with a theoretical limit of unity, which can be found in a relatively large region. We demonstrate that this directionality is broadband, spanning 10's of nms, and can be observed at a large range of $n_g$'s for one PhCW. Furthermore, we demonstrate efficient directional coupling of both electric and magnetic emitters to the PhCW, making our results relevant for quantum emitters such as magnetic molecules[31].

Our observations of high directional efficiency, combined with observations that PhCWs can be used to almost perfectly extract radiation from QDs within the slab[32,33], demonstrate the two key requirements of a deterministic spin to pathway interface. In fact, recent experiments have suggested that such an interface may be viable for systems other than QDs such as, for example, atoms [23,34,35]. Such an interface, be it for QDs, atoms, or any other emitter, would allow for the entanglement of emitter spin to photonic pathway[16], or could even be used to create quantum logic gates[15], an important step towards future on-chip quantum optics applications.

**Methods**

*Fabricating the waveguides*

The fabrication of the photonic crystal waveguide begins with a silicon-on-insulator wafer (SOITEC, $200\,\text{nm}$ Si layer/ $2\,\mu\text{m}$ $SiO_2$ buffer). We have used electron-beam lithography to generate the required patterns in a resist (350 nm of ZEP 520-A, Zeon Chemicals). The patterns were transferred directly into the silicon layer of the wafer by reactive ion etching with an $SF_6$/$CHF_3$ gas mix. After removing the remaining resist, the $SiO_2$ buffer was selectively removed from beneath the photonic crystal waveguide region with a dilute HF solution, thereby forming a silicon membrane in the area of the photonic crystal waveguide. The access waveguides that feed the photonic crystal waveguide are protected from the HF by a layer of S1818 (Shipley), which is spun on prior to the HF step.

*Calculation of the waveguide eigenmodes and dispersion*

The photonic crystal waveguide (PhCW) is a dispersion engineered waveguide, where a missing row of holes in a $200\,\mathrm{nm}$ thin silicon membrane perforated with a hexagonal pattern of holes (see Fig. 1c of the main text). Light is confined in the plane of the waveguide by the photonic band gap of the surrounding holes, and is confined to the silicon slab by total internal reflection. Our waveguide has a lattice period $a = 420\,\mathrm{nm}$ and a hole radius of $r = 110\,\mathrm{nm} = 0.262a$. The holes closest to the missing row of holes are shifted outwards by $45\,\mathrm{nm} = 0.11a$. The waveguide eigenmodes $\mathbf{e}_L(\mathbf{r},\omega)$ and eigenfrequencies are calculated with the freely available MIT Photonic Bands[36], which determines the eigenmodes of the structure using a plane-wave basis set and periodic boundary conditions. We have used a supercell of dimensions $a \times 11a\sqrt{3} \times 10h$, where $h$ is the thickness of the Si slab, which is large enough to avoid interactions between neighbouring supercells. The calculations use a grid-size of $a/16$, which ensures convergence of the eigenvalues to better than 0.1%. We used a refractive index of Si of 3.48, suitable for wavelengths around $1580\,\mathrm{nm}$.

**Acknowledgments:**

The authors thank C.I. Osorio and D. van Oosten for useful discussions. This work is supported by NanoNextNL of the Government of the Netherlands and 130 partners and part of the research program of the Stichting voor Fundamenteel Onderzoek der Materie (FOM), which is financially supported by the Nederlandse organisatie voor Wetenschappelijk Onderzoek (NWO) and part of this work has been funded by the project "SPANGL4Q", which acknowledges the financial support of the Future and Emerging Technologies (FET) programme within the Seventh Framework Programme for Research of the European Commission, under FET-Open grant number: FP7-284743.

**Competing financial interest statement:**

The authors declare no competing financial interests

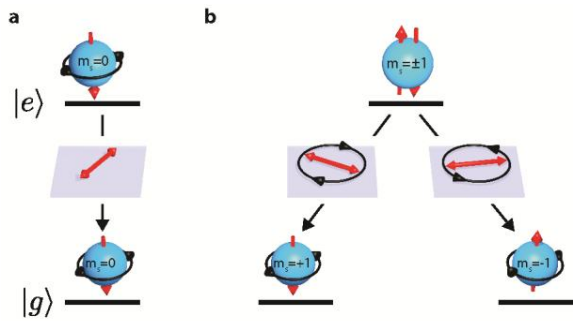

**Figure 1: Schematic of dipole mediated transitions.** Sketch of an emitter decaying from its excited state $|e\rangle$, to its ground state $|g\rangle$. The ground and the excited state electron have the same spin, for example $m_s = -1$. The transition between the two levels, a quantum process, results in the emission of a photon. This emission can be thought of as occurring due to a transition dipole (middle row), which arises due to the redistribution of charges in the emitter during the transition. **a** A transition that does not change the emitters total angular momentum. Such a transition is mediated by a linear transition dipole. **b** Sketch of an emitter prepared in an excited state that is a superposition of $m_s = -1$, $m_s = +1$, decaying with equal probability to either of the two ground states[37]. This decay is associated with a circular transition dipole, as sketched in the middle row, and the emission of a circularly polarized photon [37].

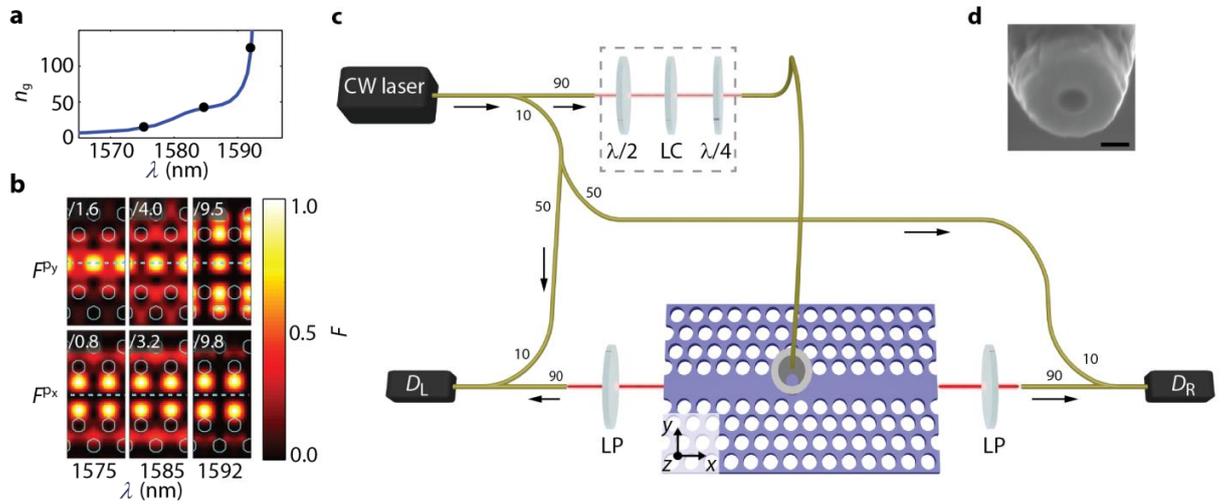

**Figure 2: Illumination-mode NSOM mimics dipolar emission into a PhCW. a** Calculated group index of the PhCW. **b** Calculated directional enhanced emission factor (for light travelling to the left) for $x-$ and $y-$ oriented electric dipoles 20 nm above the PhCW, at wavelengths marked in **a**. Blue circles indicate the edges of the 110 nm holes in the PhCW and blue dashed lines indicate the center of the waveguide. **c** Schematic of the near-field scanning optical microscope, operating in illumination mode, that acts as a dipolar source. Light from a CW continuous-wave laser is split into reference and signal branches with the ratios shown. The polarization of the light in the signal branch, and hence the orientation of the dipole mimicked by the NSOM probe, shown in **d**, is controlled by a the polarization control scheme (dashed grey rectangle) that includes a half-wave plate ($\lambda/2$), a liquid crystal variable wave plate (LC) and a quarter-wave plate ($\lambda/4$). Light that couples to the PhCW travels to the left, or right, where it is detected in $D_L$ and $D_R$, respectively, in an interferometric manner. **d** SEM image of the near-field probe used in this work, with a 200 nm scale bar.

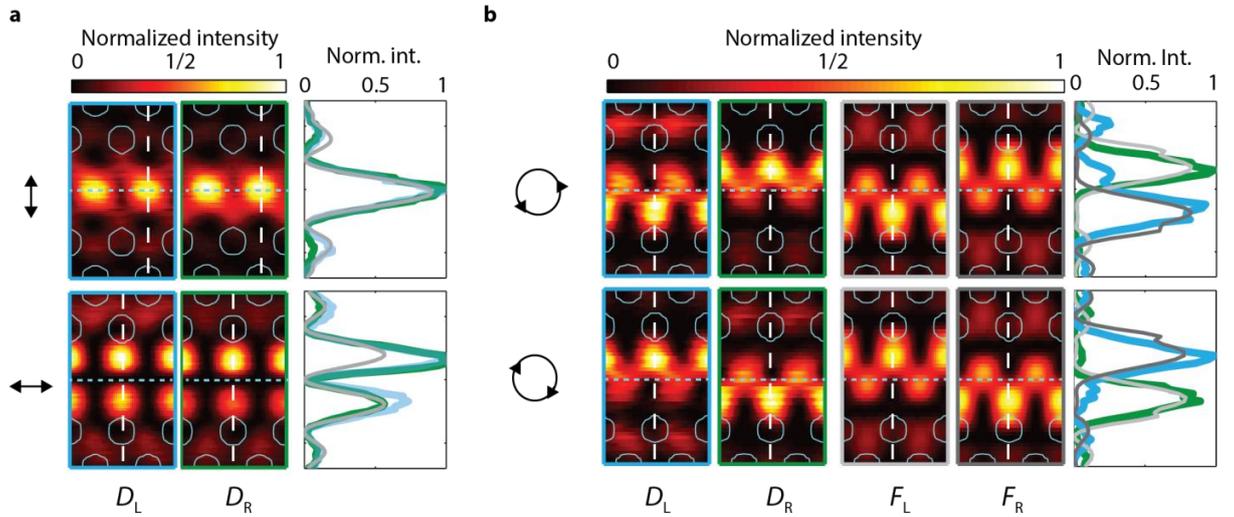

**Figure 3: Dipolar emission into a PhCW. a** Experimentally collected emission maps collected with the left ($D_L$, left column) and right ($D_R$, right column) detectors, at $1575\,\text{nm}$ with linear dipoles (orientation indicated by the black arrows). Line traces are taken along the white dashed lines shown in the corresponding row, with the blue (green) lines corresponding to line traces through $D_L$ ($D_R$). Grey lines show cuts through the calculated $F(\hat{\mathbf{p}}_x)$ (bottom panel) and $F(\hat{\mathbf{p}}_y)$ (top panel), shown in Fig. 2b. **b** Emission maps of right- (top row) and left-handed (bottom row) circularly polarized dipoles. In each row the two left (right) panels show experimental (theoretical) emission maps. As in **a**, the line traces (along the white dashed lines) correspond to cuts through $D_L$ ($D_R$), and are shown together with calculated cuts through $F_L$ (dark grey lines) and $F_R$ (light grey lines). In all maps, light blue lines show the position of the holes of the PhCW and blue dashed lines mark the center of the waveguide.

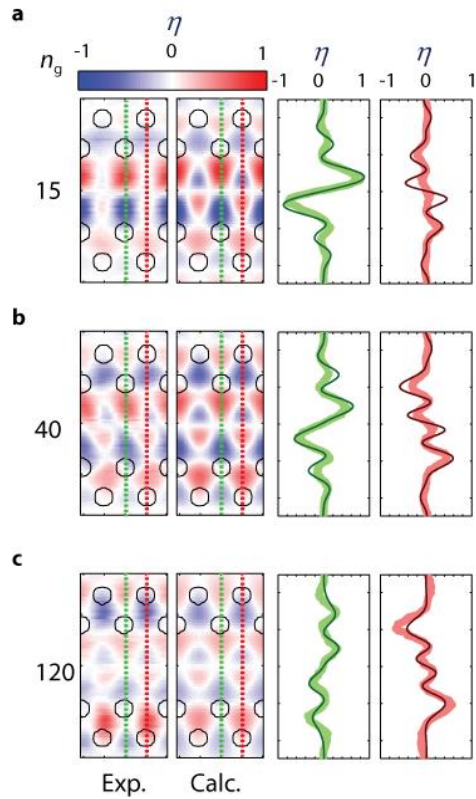

**Figure 4: Spin-to-path coupling efficiency.** Experimental and calculated $\eta$ maps for a combined **p** and **m** dipole source at **a** 1575 nm where $n_g = 15$, **b** 1585 nm where $n_g = 40$, and **c** 1592 nm where $n_g = 120$. In **a, b, c** experimental results are shown in the left panel, calculations in the middle panel, and cuts along the dashed lines in the final panels. In these cuts, experimental (theoretical) data is shown in thick (thin) curves.

# A scalable interface between solid-state and flying qubits: observations of near-unity dipole helicity to photon pathway coupling
## - SUPPLEMENTARY MATERIAL -

B. le Feber,[1] N. Rotenberg,[1] and L. Kuipers[1, *]

[1]*Center for Nanophotonics, FOM Institute AMOLF,
Science Park 104, 1098 XG, Amsterdam, The Netherlands*

(Dated: November 25, 2014)

## S1. COMPLEX DIPOLE EMISSION NEAR A PHOTONIC CRYSTAL WAVEGUIDE

The way in which an emitter radiates when placed near a nanophotonic structure can be understood in a classical framework. Specifically, the field (at $\mathbf{r}'$) of an emitter (at position $\mathbf{r}$ and frequency $\omega$) placed in an arbitrary environment can be conveniently expressed using the system's Green's function $\overleftrightarrow{\mathbf{G}}(\mathbf{r}, \mathbf{r}'; \omega)$. That is, $\overleftrightarrow{\mathbf{G}}(\mathbf{r}, \mathbf{r}; \omega)$ describes how the environment affects an emitter's ability to radiate. In a Green's function formalism [1], the enhanced emission factor compared to vacuum, is given by

$$F\left(\hat{\mathbf{d}}, \mathbf{r}, \omega\right) = \frac{\mathrm{Im}\left(\hat{\mathbf{d}}^* \cdot \overleftrightarrow{\mathbf{G}}\left(\mathbf{r}, \mathbf{r}; \omega\right) \cdot \hat{\mathbf{d}}\right)}{G_0\left(\omega\right)}, \tag{S1}$$

where $\hat{\mathbf{d}}$ represents the orientation of a classical or a transition dipole, $G_0\left(\omega\right) = \mathrm{Im}\left(\hat{\mathbf{d}}^* \cdot \overleftrightarrow{\mathbf{G}}^{\mathrm{hom}}\left(\mathbf{r}, \mathbf{r}; \omega\right) \cdot \hat{\mathbf{d}}\right)$ is the solution in a homogeneous medium, which for an electric dipole reduces to $G_0\left(\omega\right) = \omega^3\sqrt{\epsilon_d}/\left(6\pi c^3\right)$, where $\epsilon_d = \epsilon_d\left(\mathbf{r}, \omega\right)$ is the relative permittivity. In this work, to find $F$ for a circularly polarized dipole, we specifically take into account that $\hat{\mathbf{d}}$ can be complex. Assuming that the emission of a nearby emitter is completely into the photonic crystal waveguide (PhCW) its Green's function can be found analytically to be $\overleftrightarrow{\mathbf{G}}\left(\mathbf{r}, \mathbf{r}; \omega\right) = ia\omega n_g\left(\omega\right)/2c\left(\mathbf{e}_R\left(\mathbf{r}, \omega\right) \otimes \mathbf{e}_R^*\left(\mathbf{r}, \omega\right)\right)$ [1]. However, in this equation the left- and rightwards propagating modes contribute to the same Green's function. To separate the effect of the left- and rightwards propagating modes on the radiation of the emitter, we write

$$\overleftrightarrow{\mathbf{G}}_{L,R}\left(\mathbf{r}, \mathbf{r}; \omega\right) = \frac{ia\omega n_g\left(\omega\right)}{4c}\mathbf{e}_{L,R}\left(\mathbf{r}, \omega\right) \otimes \mathbf{e}_{L,R}^*\left(\mathbf{r}, \omega\right), \tag{S2}$$

where $a$ is the PhCW lattice period, $\omega$ the emission frequency, $n_g\left(\omega\right)$ the group index and $\mathbf{e}_{L,R}\left(\mathbf{r}, \omega\right)$ the modal field of the left- ($L$) and rightwards ($R$) propagating modes. To find the emission enhancement factor $F_{L,R}\left(\hat{\mathbf{p}}, \mathbf{r}, \omega\right)$ for a complex electric dipole ($\mathbf{p}$), we insert Eq. S1 into Eq. S2, which results in

$$F_{L,R}\left(\hat{\mathbf{p}}, \mathbf{r}, \omega\right) = \frac{6\pi c^2 a n_g\left(\omega\right)}{4\omega^2 \sqrt{\epsilon_d\left(\mathbf{r}, \omega\right)}} \, \mathrm{Im}\left(i\left\langle\hat{\mathbf{p}}\right| \mathbf{e}_{L,R}\left(\mathbf{r}, \omega\right) \otimes \mathbf{e}_{L,R}^*\left(\mathbf{r}, \omega\right) \left|\hat{\mathbf{p}}\right\rangle\right). \tag{S3}$$

Next we note that

$$\hat{\mathbf{p}}^* \cdot \left(\mathbf{e}_{L,R}\left(\mathbf{r}, \omega\right) \otimes \mathbf{e}_{L,R}^*\left(\mathbf{r}, \omega\right)\right) \cdot \hat{\mathbf{p}} = \left(\hat{\mathbf{p}}^* \cdot \mathbf{e}_{L,R}\left(\mathbf{r}, \omega\right)\right)\left(\hat{\mathbf{p}} \cdot \mathbf{e}_{L,R}^*\left(\mathbf{r}, \omega\right)\right) = \left|\hat{\mathbf{p}}^* \cdot \mathbf{e}_{L,R}\left(\mathbf{r}, \omega\right)\right|^2, \tag{S4}$$

and by inserting Eq. S4 into Eq. S3, we find that

$$F_{L,R}\left(\hat{\mathbf{p}}, \mathbf{r}, \omega\right) = \frac{3\pi c^2 a n_g\left(\omega\right)}{2\omega^2 \sqrt{\epsilon_d\left(\mathbf{r}, \omega\right)}} \, \mathrm{Im}\left(i\left|\hat{\mathbf{p}}^* \cdot \mathbf{e}_{L,R}\left(\mathbf{r}, \omega\right)\right|^2\right), \tag{S5}$$

which straightforwardly reduces to

$$F_{L,R}\left(\hat{\mathbf{p}}, \mathbf{r}, \omega\right) = \frac{3\pi c^2 a n_g\left(\omega\right)}{2\sqrt{\epsilon_d\left(\mathbf{r}, \omega\right)}} \left|\hat{\mathbf{p}}^* \cdot \mathbf{e}_{L,R}\left(\mathbf{r}, \omega\right)\right|^2. \tag{S6}$$

---

*Electronic address: `kuipers@amolf.nl`

In the air above the PhCW $\epsilon_d(\mathbf{r},\omega) = 1$, and Eq. S6 reduces to

$$F_{L,R}(\hat{\mathbf{p}},\mathbf{r},\omega) = \frac{3\pi c^2 a n_g(\omega)}{2\omega^2} |\hat{\mathbf{p}}^* \cdot \mathbf{e}_{L,R}(\mathbf{r},\omega)|^2, \tag{S7}$$

which is identical to equation (1) in the main text for an electric dipole

## S2. CALCULATED MAGNETIC DIPOLE EMISSION ENHANCEMENT

Our expression of the emission enhancement for an electric dipole can be used to find the expression for the emission enhancement of a magnetic dipole. Specifically, to find the emission rate enhancement factor for a magnetic dipole we perform the substitution suggested in Ref. [2]. Replacing $[\mathbf{E}, \mathbf{H}, \mu_0\mu, \epsilon_0\epsilon, \mathbf{p}]$ with $[\mathbf{H}, -\mathbf{E}, \epsilon_0\epsilon, \mu_0\mu, \mu\mathbf{m}]$, we find that for a magnetic dipole

$$F_{L,R}(\hat{\mathbf{m}},\mathbf{r},\omega) = \frac{3\pi c^2 a n_g(\omega)}{2\omega^2} |\hat{\mathbf{m}}^* \cdot \mathbf{e}_{L,R}(\mathbf{r},\omega)|^2, \tag{S8}$$

where the dot product effectively selects the magnetic fields of the mode that the dipole couples to.

Interestingly, for both polarizations and at all wavelengths, the emission enhancement of magnetic dipoles is roughly a factor 6 larger than for electric dipoles. Notably, in parallel to the similarity of the calculated electric and magnetic modal fields 20 nm above the waveguide [3], the pattern of the calculated emission enhancement factor for a magnetic dipole (see Fig. S1b) is almost identical to that for an electric dipole (see Fig. S1a) 20 nm above the waveguide, although at all wavelengths small differences are present.

In Fig. S1c we show the calculated emission enhancement factor for and electric dipole placed at the center height of the slab. We also note that due to the TE symmetry of the mode, $F_{L,R}(\hat{\mathbf{m}},\mathbf{r},\omega) = 0$, and hence magnetic dipole emission greatly depends on the distance to the slab center.

## S3. MEASURED LINEAR ELECTRIC AND MAGNETIC DIPOLE EMISSION

Fig. S2a-f (left column) show the emission maps generated with $x$- and $y$-polarized illumination. The maps measured with $x$-polarized illumination show slight differences, compared to the theoretical maps for a $p_x$ and $m_y$ (see Fig. S1). Specifically, we find that compared to the calculated maps, at 1575 nm, the profile of the emission map between maxima is differently shaped (see Fig. S2a, b, arrow 1). At 1585 nm (see Fig. S2c, d, arrows 2 and 3) there is more intensity to the side of the waveguide and at 1592 nm (see Fig. S2e, f, arrow 4) we again observe a slight suppression of the emission at the waveguide center. Furthermore, a close examination of these figures reveals that at all wavelengths

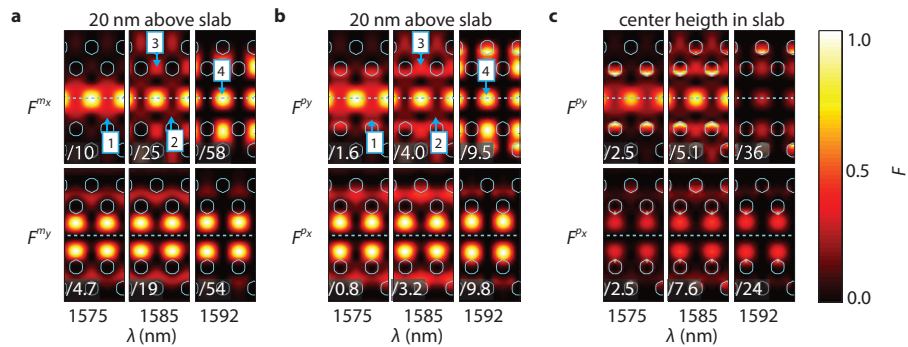

FIG. S1: **Calculated emission enhancement of linear electric and magnetic dipoles. a** Emission enhancement factor for $x$- (top row of panels) and $y$-oriented (bottom row) magnetic dipoles 20 nm above the PhCW. **b**, **c** Emission enhancement factor for $y$- (top row of panels) and $x$-oriented (bottom row) electric dipoles 20 nm above the PhCW **b** and in the center of the slab **c**. The factors in the bottom left of each panel show the factors used to normalize the maximum enhancement in each panel to one. The blue dashed line indicates the plane of mirror symmetry of the crystal, the blue circles indicate the PhCW holes. The blue labels 1-4 in **a** and **b** correspond to the positions labeled in Fig. S2.

a slight left-right asymmetry is present[1] (for example arrows 1,3 indicate positions of high asymmetry). Interestingly, an asymmetry in the emission direction of *linear* dipoles, was recently reported for the combined emission of linear electric and magnetic dipoles and this asymmetry was also used quantify the relative strength of these dipoles [4–6]. Hence, our observation of directional emission of a linear dipole suggests that we could use the emission maps measured with linear dipoles to quantify the relative electric and magnetic dipole strength of our probe.

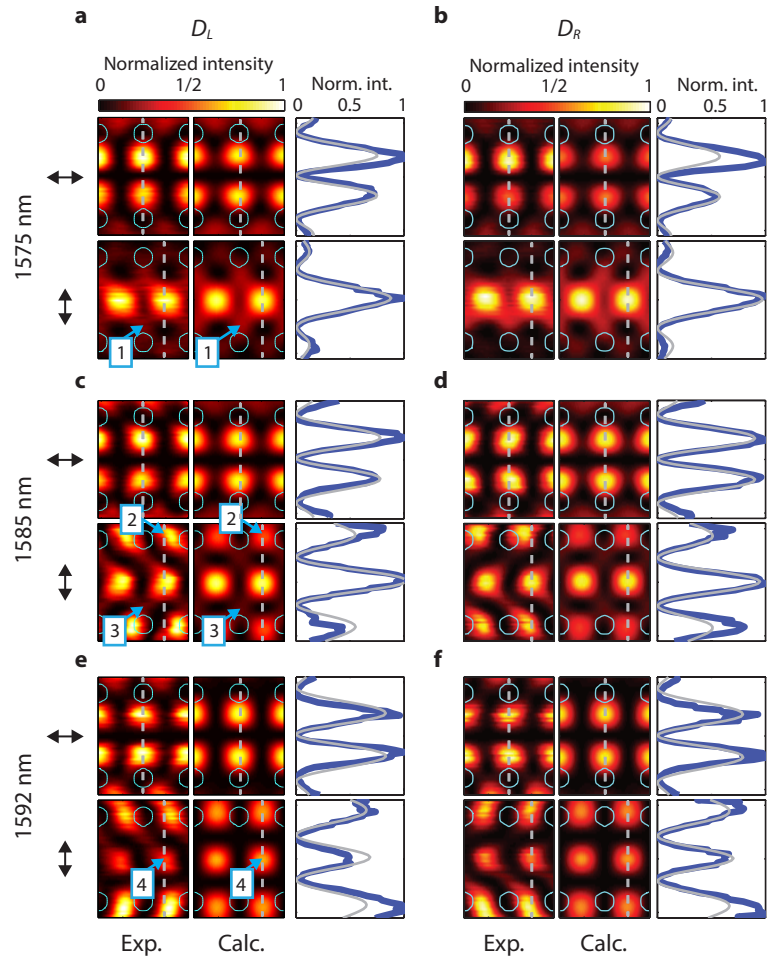

FIG. S2: **Measured linear electric and magnetic dipole emission. a**, **c**, **e** (**b**, **d**, **f**) show the calculated and measured leftwards (rightwards) emission. In **a-f**, the top (bottom) row shows $x$- ($y$-) polarized emission. The left (middle) column of panels shows the measured (calculated) emission maps and the right column of panels shows line cuts taken along the gray dashed line. In the line traces, the blue (gray) line shows the experimental (calculated) intensity. Blue labels in **a**, **c**, **e** show positions where the calculated emission of a combined electric magnetic emitter particularly improves agreement with measurements. Black arrows indicate emission polarization. Blue circles indicate the holes in the PhCW.

To quantify the contribution of both electric and magnetic dipolar emission, we approximate our aperture probe as a combined electric [7] and magnetic [8] dipole source. Additionally, compared to Ref. [3], we now measure closer to the surface of the crystal, where higher wavevectors are more abundant. Because our probe cannot couple to some of these wavevectors, we now also need to take the finite size of our tip into account. Although we could invoke the optical reciprocity theorem to predict our measurements (see Ref. [3]), this would not provide extra experimental insight and the spatially different profiles of the electric and magnetic reciprocal fields make assigning a single strength to electric and magnetic dipole emission non-trivial. Hence, we fit our experimental measurements with a superposition of the electric and magnetic modal fields that we convoluted with a (210 nm diameter) disk. Specifically, we approximate

---

[1] We stress that the directional emission of these linear dipoles is completely unrelated to the helicity of the emitting dipoles, which do not have a defined helicity because they are linear. Hence, this directional emission cannot be coupled to a helicity or a spin transition.

the measured signal with

$$F_{L,R}\left(\mathbf{p}_x, \mathbf{m}_y; \mathbf{r}, \omega\right) = \left|\alpha E_x^{con}\left(\mathbf{r}, \omega\right) - \beta Z_0 H_y^{con}\left(\mathbf{r}, \omega\right)\right|^2, \tag{S9a}$$

$$F_{L,R}\left(\mathbf{p}_y, \mathbf{m}_x; \mathbf{r}, \omega\right) = \left|\alpha E_y^{con}\left(\mathbf{r}, \omega\right) + \beta Z_0 H_x^{con}\left(\mathbf{r}, \omega\right)\right|^2, \tag{S9b}$$

where $Z_0$ is the impedance of free space, $E_{x,y}^{con}(\mathbf{r}, \omega)$ and $H_{y,x}^{con}(\mathbf{r}, \omega)$ are the convoluted electric and magnetic modal field components and $\alpha$ and $\beta$ are complex fitting parameters that quantify the relative electric and magnetic coupling strengths. As sketched in Fig. S3, rotating the injection polarization causes a sign change between the in-plane electric and magnetic fields. Therefore, Eqs. S9a and b differ by a minus sign that ensures a consistent phase between the electric and magnetic coupling for both injection polarizations.

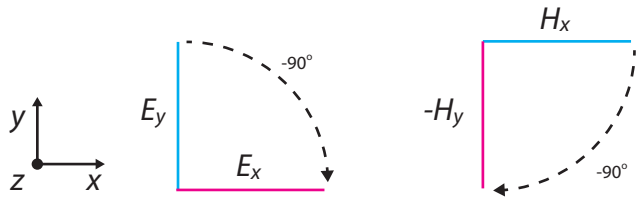

FIG. S3: **Rotating the illumination polarization.** (left) Coordinate system. (middle, right) Blue (pink) lines show the field orientation for $x$- ($y$-) polarized illumination, respectively. Rotating the illumination polarization -90 degrees changes $E_y$ to $E_x$ (middle) and $H_x$ to $H_y$ (right).

The fits that are obtained using Eqs. S9a and b (middle panels of Fig. S2a-f) show good agreement with the measured fields when $|Z_0\beta|/|\alpha| = 0.9$ and $\phi_\beta - \phi_\alpha = -0.3\pi$ rad. This agreement is emphasized by the line traces shown in the right panels of Fig. S2a-f. Nonetheless, the emission along the waveguide center at 1592 nm remains somewhat suppressed compared to the calculations and at 1585 nm we observe slightly lower emission efficiency to the side of the waveguide. These differences could be explained by the increased interaction between the probe and the sample at these higher $n_g$'s.

## S4. CONTROL OVER EMISSION POLARIZATION

### A. Achieving circular polarization

One of the benefits of using a NSOM to mimic the emission of a dipolar emitter is that this setup allows a high degree of the control over the dipole orientation. Here, we use the $\lambda/2$, $\lambda/4$ and liquid crystal (LC) wave plates shown in Fig. 2c (main text) to control the orientation of the dipoles that we use to approximate our NSOM tip. The LC-plate (Thorlabs LCC1113-C) that has a voltage controllable phase difference between its birefringent axes, is only used in experiments where we mimic a circular dipole.

We start by orienting our half- and quarter-wave plates analogue to the procedure used in collection mode to achieve linear ($x$-) polarization beneath the probe apex. In this situation (black dot on the Poincaré spare sketched in Fig. S4a), we know that before the $\lambda/4$, the polarization of the light is aligned with one of the birefringent axes of the combined $\lambda/4$ and fiber system. Therefore, to measure the emission of a $y$-oriented dipole, we rotate the $\lambda/2$ by 45° to align our polarization to the other axis of the combined system (moving along the green circle to VP, Fig. S4a).

We now set out to show that we can control the orientation of our dipole source to create circularly (and elliptically) polarized dipoles. To mimic a circularly polarized dipole we firstly create equal amplitude $x$- and $y$-components of the probe dipole and subsequently ensure that they oscillate in quadrature. Equal amplitude in-plane dipoles are straightforwardly achieved by rotating the $\lambda/2$ by 22.5° (move along gray dashed arrow to the gray dot, Fig. S4a). However, at this point in the experiment the birefringence of the combined fiber and $\lambda/4$ system is unknown. That is, the position of dipole orientation on the purple circle drawn in Fig. S4a is unknown at this at this point. Consequently, to achieve circular polarization we need to scan the phase between the $x$- and $y$-oriented dipoles to ensure circular polarization. To achieve circular dipole orientations, we vary the phase between the equal amplitude $x$- and $y$-components of the probe dipole using the LC-plate. Specifically, after we ensure that the birefringent axes of this waveplate are aligned to the axes of the combined fiber and $\lambda/4$ system, we scan the phase difference between the two orthogonal components of the probe dipole with the LC-plate. In this manner the dipole orientation moves along the purple line on the Poincaré sphere sketched in Fig. S4a.

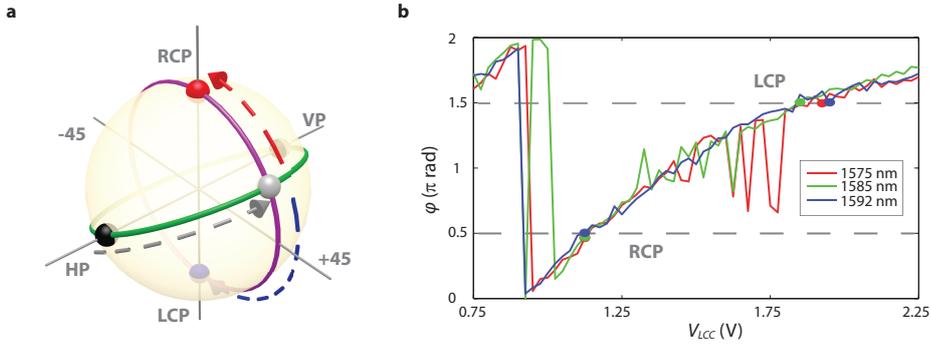

FIG. S4: **Achieving circular dipole orientation. a** Poincaré sphere, with annotated orientations of the dipoles that emits into the PhCW. Intersections between the sphere and the gray axes are labeled with the corresponding dipole orientations: linear polarized $x$, $y$ (HP,VP), linearly diagonally polarized (+45,-45), and left- and righthanded circular polarized (LCP,RCP). The green (and purple) line indicates the effect of rotating the $\lambda/2$ (and scanning the voltage on the LC) plate. **b** Relation between voltage from our liquid crystal controller ($V_{LCC}$) and the resulting phase ($\phi$) described in the text.

For each voltage ($V_{LCC}$) that we apply to the LC-plate, we find the phase difference ($\phi$) between the horizontally $[HP(\mathbf{r},\omega) = \alpha E_x^{con}(\mathbf{r},\omega) - \beta Z_0 H_y^{con}(\mathbf{r},\omega)]$ and vertically polarized dipole injection $[VP(\mathbf{r},\omega) = \alpha E_y^{con}(\mathbf{r},\omega) + \beta Z_0 H_x^{con}(\mathbf{r},\omega)]$ by fitting superpositions

$$F_{L,R}^{\mathbf{p},\mathbf{m}}(V_{LCC},\mathbf{r},\omega) = \left| HP_{L,R}(\mathbf{r},\omega) + e^{i\phi(V_{LCC})} VP_{L,R}(\mathbf{r},\omega) \right|^2 \tag{S10}$$

to our experimental emission maps on the left and right detector at that recorded at $V_{LCC}$, using the $\alpha$ and $\beta$ that we found in Sec. S3. Fig. S4b shows that the obtained via this approach show excellent agreement between the three wavelengths (1575, 1585 and 1592 nm). Furthermore, we find that we can sample an entire $2\pi$ phase difference. Having established this calibration, our approach is straightforwardly extended to mimic any complex dipole orientation. To measure the emission of circularly polarized dipoles, we pick the values of $\phi$ closest to $\pi/2$ and $3\pi/2$ for each wavelength (move along blue and red dashed arrows to blue and red spheres in Fig. S4b).

### B. Measured circular electric and magnetic dipole emission maps

We investigate the tunability of electric and magnetic dipole helicity-to-path coupling in PhCWs by injecting circularly polarized light at 1575, 1585 and 1592 nm. In Sec. S4A we give an explanation of how we ensure circularly polarized illumination. The emission maps with circularly polarized light at all wavelengths (1575, 1585 and 1592 nm, shown in Fig. S5) reveal a clear left-right asymmetry in the emission direction. Furthermore, for all wavelengths, flipping either the direction of the detector path (i.e. monitoring either $D_L$ or $D_R$) or dipole helicity results in a mirroring of the emission enhancement map about the middle of the waveguide ($y=0$, dashed blue lines in Fig. S5). This asymmetry, which is indicative of helicity-to-path coupling, is excellently reproduced by the calculations for circularly polarized emission with a combined electric and magnetic emitter.

The helicity dependence of the emission directionality is further highlighted by line traces across the waveguide (taken along brown and pink dashed lines in the emission maps, and shown by blue and green lines in the right two panels of Fig. S5). These line traces, which are in excellent agreement with the calculations for all group indices, show that maxima for one helicity are turned into minima with a helicity reversal. Additionally, these line traces highlight that the positions at which we observe the left-right asymmetry shift between the shortest and the longest wavelengths (from the brown to the pink dashed line in Fig. S5).

From these measured and calculated fields for a combined electric and magnetic emitter, we extract the $\eta$, shown in Fig. 4 of the main text, according to equation (2) of the main text. Additionally, we show the helicity-to-path coupling efficiencies of purely electric and magnetic emitters in Fig. S6. As was the case for the combined emitter, we observe that the $|\eta|_{max}$ of the individual dipoles decreases as a function of $n_g$, and that the locations of maximal emission and maximal directionality diverge. Moreover, we observe that the $\eta_{max}$ associated with $\mathbf{p}$ and $\mathbf{m}$ are no longer the same; at 1585 nm where $n_g = 40$, $|\eta|_{max}^{\mathbf{p}} = 0.8$ and $|\eta|_{max}^{\mathbf{m}} = 0.8$, while at 1592 nm where $n_g = 120$, $|\eta|_{max}^{\mathbf{p}} = 0.6$ and $|\eta|_{max}^{\mathbf{m}} = 0.5$. Furthermore, the positions where we find $|\eta|_{max}$ are different for $\mathbf{p}$ and $\mathbf{m}$, suggesting that a PhCW can interact electric and magnetic dipoles in different ways.

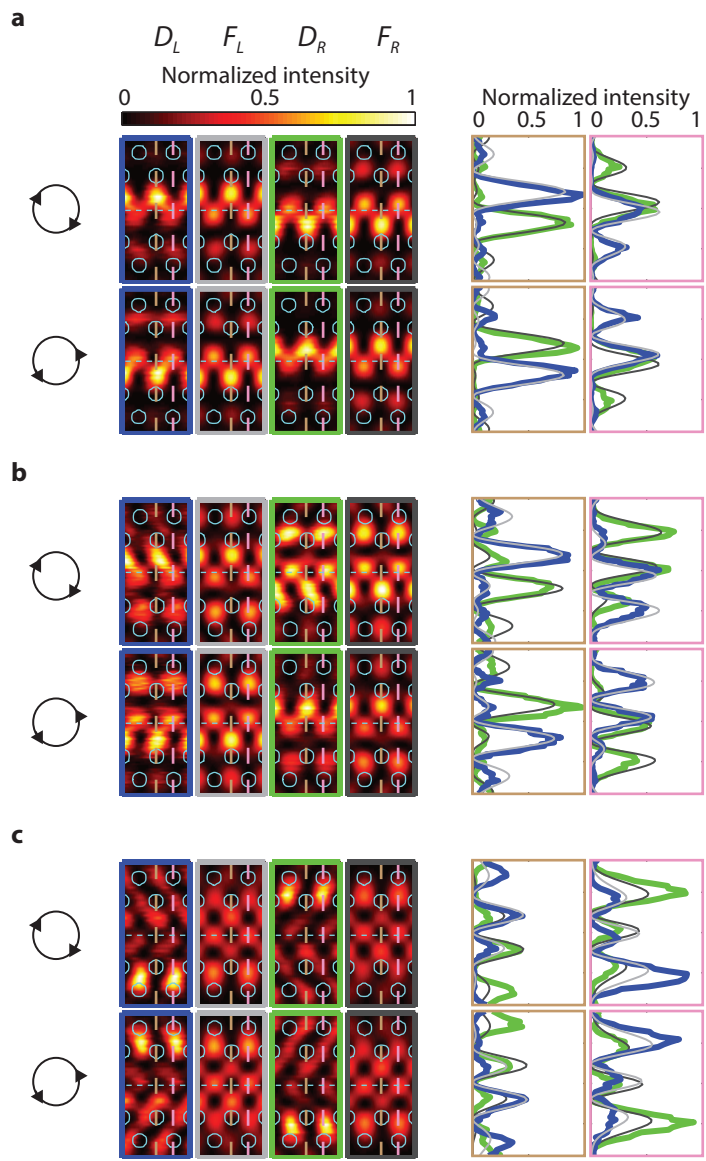

FIG. S5: **Circular dipole emission into the PhCW.** Experimental ($D_R$ and $D_L$) and calculated ($F_L$ and $F_R$) emission maps for circularly polarized emission at **a** (1575 nm), **b** (1585 nm) and **c** (1592 nm). The right two figure columns show line traces through the calculated and measured maps, along the dashed pink (right figures with line traces) and yellow (left figures with line traces) lines. Blue (green) lines show cuts through $D_L$ ($D_R$). Light (dark) gray lines show cuts through $F_L$ ($F_R$). Black arrows indicate the circular dipole handedness.

## S5. EMISSION DIRECTIONALITY

The coupling efficiency $\eta$ (main text, equation (2)) includes not only the coupling between helicity and optical path, but also the emission strength relative to the maximal emission intensity for a circularly polarized dipole. These two quantities are of themselves very interesting. For example, if $|\eta| < 1$, a structure might enable high helicity-to-path coupling, but at a relatively low emission rate. Additionally, placement inaccuracy [9] might lead to $|\eta| < 1$, but helicity can still be deterministically coupled to path.

To quantify the coupling strength between helicity and path we compute the directionality of the emission from the experimental emission maps. That is, we define the directionality, $\eta_{dir}$, as follows:

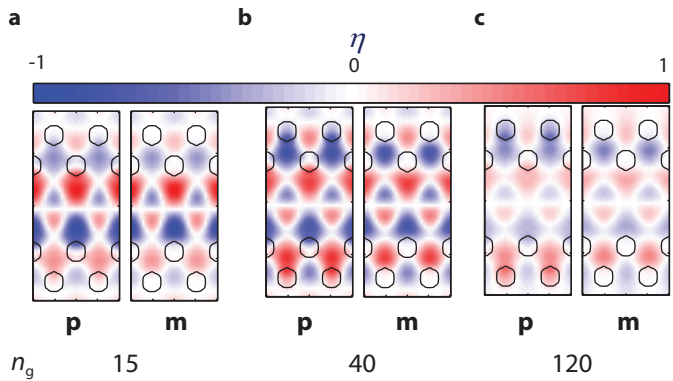

FIG. S6: **Calculated $\eta$ for electric and magnetic dipoles. a-c** Show the calculated $\eta$ for **p** and **m** dipoles emitting at **a** 1575 nm ($n_g = 15$), **b** 1585 nm ($n_g = 40$) and **c** 1592 nm ($n_g = 120$). Black contours indicate the PhCW holes.

$$\eta_{dir}(\mathbf{r},\omega) = \frac{D_L\left(\hat{\mathbf{d}}_{LCP}\right) - D_R\left(\hat{\mathbf{d}}_{LCP}\right)}{D_L\left(\hat{\mathbf{d}}_{LCP}\right) + D_R\left(\hat{\mathbf{d}}_{LCP}\right)} - \frac{D_L\left(\hat{\mathbf{d}}_{RCP}\right) - D_R\left(\hat{\mathbf{d}}_{RCP}\right)}{D_L\left(\hat{\mathbf{d}}_{RCP}\right) + D_R\left(\hat{\mathbf{d}}_{RCP}\right)}, \quad (S11)$$

where $LCP$ and $RCP$ indicate left- and right-handed dipoles, respectively.

In Fig. S7a, d, g (left panels) we show the experimental $\eta_{dir}^{exp}$ maps obtained using Eq. S11 at 1575, 1585 and 1592 nm. These maps reveal that for all wavelengths large areas of high $|\eta_{dir}|$ are available. Furthermore, these directionality maps are almost identical to the maps calculated for the combined electric and magnetic dipole emission (second column of panels from the left, Fig. S7a, d, g). The line traces across the waveguide (right two columns of panels, Fig. S7a, d, g) both confirm this excellent agreement and highlight the size of the areas where $|\eta_{dir}|^{exp} > 0.8$ (grey regions in line traces, Fig. S7a, d, g).

To gain more insight in the electric and magnetic dipole emission that gives rise to the maps in Fig. S7a, d, g, we show the separate emission from these dipoles in Fig. S7b, e, h. The $|\eta_{dir}|^{\mathbf{p,m}}$ maps for both an electric and a magnetic dipole emitter show large areas of efficient coupling. The near-unity values of $|\eta_{dir}|^{\mathbf{p,m}}$ that we find in Fig. S7b, e, h highlight that emitter helicity and photon path can be deterministically coupled on a PhCW waveguide. Furthermore, we find that although at 1585 and 1592 nm, where the maximal $|\eta|^{\mathbf{p,m}}$ is not close to one (main text), it can be that $|\eta_{dir}|$ is still near unity. Therefore, although emission into the PhCW does not equal the maximal emission of a circularly polarized emitter at these wavelengths, helicity and path are still highly coupled.

A comparison between the contours of $|\eta|$ and $|\eta_{dir}|$ in Fig. S7c, f, i reveals that at all wavelengths the contours of $|\eta_{dir}|$ enclose much larger areas then those for $|\eta|$. Hence, although emission might not equal the maximal emission at higher $n_g$, helicity and path are still strongly coupled.

## S6.   GEOMETRIC CONTROL OF COUPLING STRENGTH

To explore the possibility of achieving high $|\eta(\omega)|$ and $n_g$ simultaneously we calculate the maximally available $|\eta(\omega)|$ for a larger wavelength range and investigate the effect of a slight change in the PhCW on $|\eta(\omega)|_{max}$. We begin by computing the group index and $|\eta(\omega)|_{max}$ for each calculated PhCW mode (red lines, Fig. S8a, b). These calculations reveal that the PhCW offers near-unity helicity-to-path coupling for both electric and magnetic dipoles at wavelengths shorter than 1575 nm. Interestingly, at longer wavelengths $|\eta(\omega)|_{max}$ first drops, then recovers, before dropping near the PhCW mode-gap [with the magnetic dipole (thin lines, Fig. S8b) dropping slightly faster than the electric dipole (thick lines, Fig. S8b)]. This drop to zero for both emitter types can be understood, because, at the mode gap, the PhCW mode is a standing wave [10]. Due to the structural symmetry this standing wave is equally left- and rightwards propagating and hence cannot couple helicity to path. A precise investigation of how the maximal coupling strength and position evolve with wavelength remains interesting for future studies.

To obtain an idea of the effect of the PhCW geometry on $|\eta(\omega)|_{max}$, we calculate $|\eta(\omega)|_{max}$ for a slightly different PhCW geometry. That is, we repeat the calculations with a 220 nm thick slab (corresponding to a 20 nm increase in thickness) and 120 nm radius (a 10 nm increase) holes (blue lines Fig. S8a, b), while keeping the positions of the holes fixed. For this geometry we observe a shift of the band structure towards larger wavelengths (blue line, Fig. S8a).

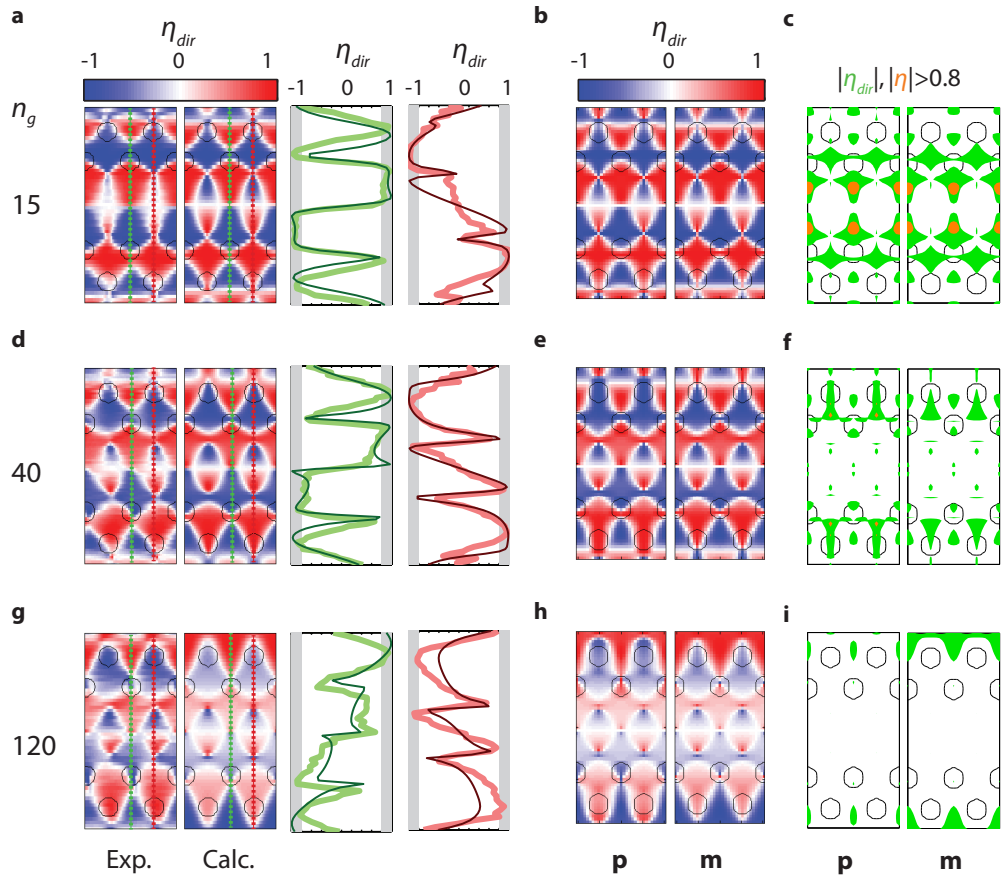

FIG. S7: **Emission directionality**. Experimental and calculated $\eta_{dir}$ maps for the combined (separate) **p** and **m** dipole sources at **a** (**b**) 1575 nm where $n_g = 15$, **d** (**e**) 1585 nm where $n_g = 40$, and **g** (**h**) 1592 nm where $n_g = 120$. In **a**, **d**, and **g**, experimental results are shown in the left panel, calculations in the middle panel, and cuts along the dashed lines in the final panels. In these cuts, experimental (theoretical) maps are shown in thick (thin) curves and grey regions indicate $|\eta| > 0.8$. In **b**, **e**, **h**, $\eta_{dir}$ maps associated with **p** (**m**) are shown on the left (right). **c**, **f**, **i** show orange (green) regions indicating where $|\eta| > 0.8$ ($|\eta_{dir}| > 0.8$) at wavelengths of 1575 nm, 1585 nm and 1592 nm, respectively. Black circles show contours of the holes in the PhCW.

Consequently, we expect that this structure enables higher $|\eta(\omega)|_{max}$ at longer wavelengths. Figure S8b shows that, indeed, the tuned PhCW geometry allows for larger $|\eta(\omega)|_{max}$, for both electric and magnetic dipoles.

## S7. CIRCULAR DIPOLE EMISSION IN THE PHCW SLAB

In Fig. S9a we show the calculated emission enhancement distributions for circular electric dipoles placed in the center of the PhCW slab. In the waveguide we also observe a strong helicity-to-path coupling efficiency as shown in Fig. S9b. Specifically, at the center of the slab we find a maximal $|\eta| > 0.99$

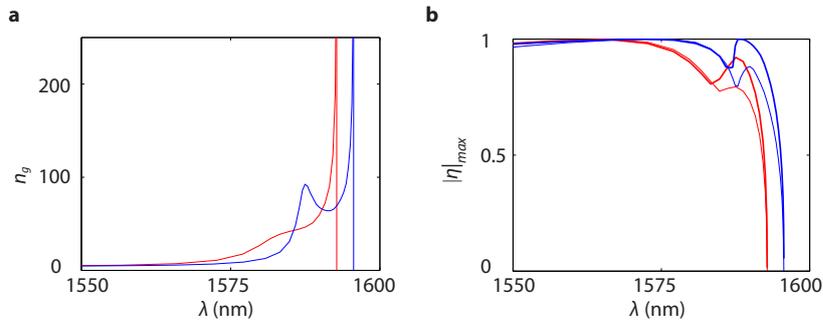

FIG. S8: **Coupling efficiency as a function of wavelength.** **a** Dispersion relation of the PhCW modes with the original (blue line) and tuned geometry (red line). **b** Coupling strength as a function of wavelength for emission with an electric (thick lines) and with a magnetic (thin lines) for the PhCW with the changed (blue lines) and with the experimental geometry (red lines).

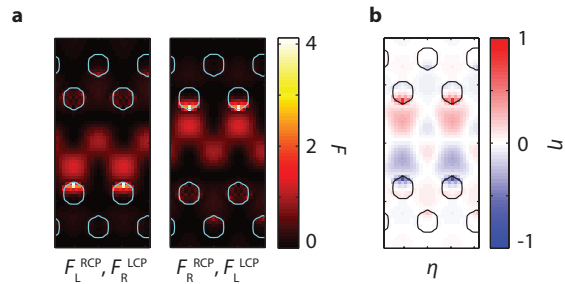

FIG. S9: **Circular dipole emission enhancement and directionality in the PhCW slab.** **a** Calculated emission enhancement factor ($F$) for circular dipoles (emitting at a free space wavelength of 1575 nm) placed in the center of the photonic crystal slab. The emission direction and handedness are indicated in the subscripts of each panel. Note that $F_L^{LCP}$ and $F_R^{RCP}$ are identical, as are $F_R^{LCP}$ and $F_L^{RCP}$. **b** Calculated helicity-to-path coupling efficiency ($\eta$) for the circular dipole emission shown in **a**. The blue circles in **a** and the black circles in **b** indicate the air holes in the silicon PhCW.